
\input harvmac
\overfullrule=0pt
\abovedisplayskip=12pt plus 3pt minus 3pt
\belowdisplayskip=12pt plus 3pt minus 3pt
%

\def\bar{\overline}
\def\np#1#2#3{Nucl. Phys. {\bf B#1} (#2) #3}
\def\pl#1#2#3{Phys. Lett. {\bf #1B} (#2) #3}

\def\cmp#1#2#3{Comm. Math. Phys. {\bf #1} (#2) #3}
\def\mpl#1#2#3{Mod. Phys. Lett. {\bf #1} (#2) #3}
\font\zfont = cmss10 
\font\litfont = cmr6

\def\bigone{\hbox{1\kern -.23em {\rm l}}}
\def\ZZ{\hbox{\zfont Z\kern-.4emZ}}
\def\half{{\litfont {1 \over 2}}}

\def\mg1{{\cal M}_{g,1}}
\def\mgn{{\cal M}_{g,n}}
\def\cmg1{{\overline{\cal M}}_{g,1}}

\def\tbar{{\bar t}}
\def\cF{{\cal F}}
\def\an{\alpha_n}
\def\amn{\alpha_{-n}}
\def\psibar{{\bar\psi}}
\def\Mhat{{\hat M}}
\def\mhat{{\hat m}}


\def\np{Nucl. Phys.}
\def\pl{Phys. Lett.}

\def\cmp{ Commun. Math. Phys.}
\def\ijmp{Int. J. Mod. Phys.}
\def\mpl{Mod. Phys. Lett.}

\def\pr{Phys. Rev.}

\def\KMINUS{E. Witten, \np\ {\bf B371} (1992) 191\semi
S. Mukhi and C. Vafa, hep-th/9301083, \np\ {\bf B407} (1993) 667\semi
N. Ohta and H. Suzuki, hep-th/9310180, \mpl\ {\bf A9} (1994) 541\semi
D. Ghoshal and S. Mukhi, hep-th/9312189, \np\ {\bf B425} (1994) 173\semi
A. Hanany, Y. Oz and R. Plesser, hep-th/9401030, \np\ {\bf B425}
(1994) 150\semi
Y. Lavi, Y. Oz and J. Sonnenschein, hep-th/9406056, \np\ {\bf
B431} (1994) 223\semi
D. Ghoshal, C. Imbimbo and S. Mukhi, hep-th/9410034, \np\ {\bf B440}
(1995) 355.}
\def\DMP{R. Dijkgraaf, G. Moore and R. Plesser, hep-th/9208031,
\np\ {\bf B394} (1993) 356.}

\def\MOORETC{G. Moore, \np\ {\bf B368} (1992) 557 \semi
G. Moore and R. Plesser, hep-th/9203060, \pr\ {\bf D46} (1992) 1730\semi
G. Moore, R. Plesser and S. Ramgoolam, hep-th/9111035, \np\ {\bf B377} (1992)
143.}

\def\KONTS{M. Kontsevich, \cmp\ {\bf 147} (1992) 1.}
\def\GENKONTS{M. Adler and P. van Moerbeke, \cmp\ {\bf 147} (1992)
24\semi
S. Kharchev, A. Marshakov, A. Mironov, A. Morozov and A. Zabrodin,
\hfil\break hep-th/9111037, \pl\ {\bf B275} (1991) 311.}
\def\MOREKONTS{C. Itzykson and J.B. Zuber, hep-th/9201001,
\ijmp\ {\bf A7} (1992) 5661\semi
P. Di Francesco, C. Itzykson and J.B. Zuber, hep-th/9206090,
\cmp\ {\bf 151} (1993) 193.}
\def\VIRCON{E. Witten, IASSNS-HEP-91-24, in ``Proceedings,
Differential geometric methods in theoretical physics'' Vol. 1, p.176,
New York (1991)\semi
Yu. Makeenko and G.W. Semenoff, \mpl\ {\bf A6} (1991) 3455\semi
D.J. Gross and M.J. Newman, \np\ {\bf B380} (1992) 1992.}
\def\PENNERETC{R. Penner, \cmp\ {\bf 113} (1987) 299\semi
R. Penner, J. Diff. Geom. 27 (1988) 35.}
\def\DISTVAF{J. Distler and C. Vafa, \mpl\ {\bf A6} (1991) 259.}
\def\WINF{J. Avan and A. Jevicki, \pl\ {\bf B266} (1991)
35; \pl\ {\bf B272} (1991) 17\semi
D. Minic, J. Polchinski and Z. Yang, \np\ {\bf B369} (1991) 324\semi
S.R. Das, A. Dhar, G. Mandal and S.R. Wadia, hep-th/9110021, \ijmp\ {\bf
A7} (1992) 5165.}
\def\KAZAKOV{V.A. Kazakov, M. Staudacher and T. Wynter,
hep-th/9502132, LPTENS-95/9.}
\def\VIR{R. Dijkgraaf, H. Verlinde and E. Verlinde, \np\ {\bf B348}
(1991) 435\semi
M. Fukuma, H. Kawai and R. Nakayama, \ijmp\ {\bf A6} (1991) 1385.}
\def\CHEKMAK{L. Chekhov and Yu. Makeenko, hep-th/9202006,
 \pl\ {\bf B278} (1992) 271\semi
L. Chekhov and Yu. Makeenko, hep-th/9201033, \mpl\ {\bf A7} (1992) 1223.}
\def\BCOV{M. Bershadsky, S. Cecotti, H. Ooguri and C. Vafa,
hep-th/9302103, \np\ {\bf B405} (1993) 279.}
\def\WITTENBK{E. Witten, hep-th/9306122, published in Salamfest 1993. }
\def\DIJK{R. Dijkgraaf, hep-th/9201003,
published in NATO ASI, Carg\`ese, 1991.}
\def\IM{C. Imbimbo and S. Mukhi, \np\ {\bf B364} (1991) 662.}

{\nopagenumbers
\Title{\vtop{\hbox{hep-th/9505127}
\hbox{CERN-TH/95-126}
\hbox{TIFR/TH/95-23}}}
{\centerline{The Topological Matrix Model of $c=1$ String}}
\ \vskip -1truecm
\centerline{Camillo Imbimbo\foot{E-mail: imbimbo@vxcern.cern.ch}}
\vskip 4pt
\centerline{{\it Theory Division, CERN, CH-1211 Geneva 23,
Switzerland}\foot{On leave from
{\it INFN, Sezione di Genova, Genoa, Italy}}}
\vskip 7pt
\centerline{Sunil Mukhi\foot{E-mail: mukhi@theory.tifr.res.in}}
\vskip 4pt
\centerline{\it Theory Division, CERN, CH-1211 Geneva 23, Switzerland}
\centerline{and}
\centerline{\it Tata Institute of Fundamental Research,}
\centerline{\it Homi Bhabha Rd, Bombay 400 005,
India\foot{Permanent address}}
\ \medskip
\centerline{ABSTRACT}

We derive a Kontsevich-type matrix model for the $c=1$ string directly
from the $W_\infty$ solution of the theory. The model that we obtain
is different from previous proposals, which are proven to be
incorrect. Our matrix model contains the Penner and Kontsevich cases,
and we study its quantum effective action. The simplicity of our model
leads to an encouraging interpretation in the context of
background-independent non-critical string field theory.

\ \vfill
\ifx\answ\bigans
\leftline{CERN-TH/95-126}
\leftline{TIFR/TH/95-23}
\leftline{May 1995}
\else\leftline{May 1995}\fi
\eject}
\ftno=0
\newsec{Introduction}
For $c<1$ non-critical string theories, there exists a remarkable
description of the generating function in terms of one-matrix models
of Kontsevich
type\ref\konts{\KONTS}\ref\genkonts{\GENKONTS}\ref\morekonts{\MOREKONTS}.
These give rise to a graphical expansion which corresponds to the
``fat-graph'' cell-decomposition of the moduli spaces $\mgn$ of
Riemann surfaces\ref\penneretc{\PENNERETC}.  These models play an
important role in understanding topological gravity and hence, one
hopes, the fundamental formulation of string theory.

One way to infer the equivalence of the $c<1$ string theory and the
corresponding Kontsevich-type models is to show that the latter
satisfy the Virasoro and $W_n$ constraints\ref\vircon{\VIRCON}\
which encode the complete perturbative solution of the
former\ref\vir{\VIR}.

The solution of $c=1$ string theory is given by $W_\infty$
identities\ref\winf{\WINF}. For the compactified theory at the
self-dual radius, the singlet sector perturbation series is neatly
encoded in the $W_\infty$ recursion relations obtained by Dijkgraaf,
Moore and Plesser\ref\dmp{\DMP}\ directly from matrix quantum
mechanics via a coherent-state representation of the tachyon
scattering process\ref\mooretc{\MOORETC}. In this paper we will show
that this solution can be used to directly derive a matrix-model of
Kontsevich type, for the $c=1$ string at the self-dual radius.

A Kontsevich-type matrix model for $c=1$ (which we refer to below as the
DMP model) was actually presented in Ref.\dmp, where it was obtained
again from the coherent-state description. The model that we derive
below differs from theirs in several distinctive ways. We examine very
carefully the DMP model and show that it is incorrect -- their model
does not in fact satisfy the $W_\infty$ constraints. We demonstrate
this first by general arguments, and then by explicit computation of
some tachyon correlators. Next we identify an error in their paper,
after correcting which we indeed recover our model.

An earlier proposal for a Kontsevich-type model to describe $c=1$
string theory, due to Chekhov and Makeenko\ref\chekmak{\CHEKMAK}\
turns out to be somewhat closer to the correct model that we present
here, although it is not quite right either.

The matrix model that we obtain is astonishingly simple, and we
analyse it in some detail. In Section 2 we derive the model from
$W_\infty$. In Section 3 we compare it with previous proposals and
show that the latter do not describe $c=1$ string theory. In Section
4 we give an alternative derivation of our model following the
technique of Refs.\ref\dijk{\DIJK},\dmp. In Section 5 we examine the
relation of our model to the Penner matrix model describing the Euler
characteristic of punctured surfaces, and to the original Kontsevich
model describing intersection theory on moduli space. In Section 6 we
derive the quantum effective action and show that it generates $c=1$
string amplitudes via tree graphs. In Section 7 we observe that our
Kontsevich-type model can be thought of as a background-independent
string field theory coupled to an external source. Expanding about
various possible minima then leads to the $c=1$ and $c<1$ strings, the
latter arising from condensation of particular negative-momentum
tachyons.

\newsec{The $c=1$ Kontsevich Model from $W_\infty$}
The tachyon operators $T_n$ of 2D string theory compactified on a circle
of unit radius are labelled by the integer-valued momentum $n$.
It is convenient to introduce an infinite number of variables
$t_n$ with $n=1,2,\ldots,$ in correspondence with the
tachyons of positive momentum $n$, and analogous variables $\tbar_n$
for the negative-momentum tachyons. The generating functional for
the correlation functions of all tachyons operators will be a function
$Z (t, \tbar)$ of the $t_n$ and $\tbar_n$.

The $W_\infty$ solution\dmp\ of this string theory is encoded in the
following recursion relation:
\eqn\wconstraints{
{1\over \mu^2}{
\partial Z_{W_\infty}\over \partial
\tbar_n}(t,\tbar) =
{1\over (n+1)}\oint dz\, :\!{\rm e}^{-i\mu \phi (z)} \Big({
\partial_z\over i\mu}\Big)^{n+1}{\rm e}^{i\mu \phi (z)}\! :\,
Z_{W_\infty},}
where the bosonic field $\phi(z)$ is the following operator:
\eqn\potential{\partial \phi(z) = {1\over z} + \sum_{n>0} n t_n
z^{n-1}-{1\over \mu^2} \sum_{n>0} {\partial\over \partial t_n}
z^{-n-1}.}

We now show that one can in fact construct a matrix model with
logarithmic potential starting directly from the above expression.
First, change variables in the $W_\infty$ relation using the
Frobenius-Miwa-Kontsevich transformation:
\eqn\frob{
i\mu\, t_n = -{1\over n} \tr
A^{-n}, }
where $A$ is a fixed $N\times N$ Hermitian matrix. In the limit of large
$N$, all the couplings become independent, but at finite $N$ everything
continues to be valid in a subspace of the parameter space where only
the first $N$ $t_k$ are independent. One finds
\eqn\wwitha{
\eqalign{
{1\over \mu^2}{
\partial Z_{W_\infty}\over \partial
\tbar_n}(t,\tbar) &=
{1\over n+1}\oint dz\, :\!{z^{-i\mu}{\rm e}^{{1\over i\mu}\sum_{k>0}
{z^{-k}\over k}{\partial\over\partial t_k}}
\over \det\,(1-zA^{-1})}\cr
&\qquad\times \Big({\partial_z\over i\mu}\Big)^{n+1} z^{i\mu}\,
{\rm e}^{-{1\over i\mu}\sum_{k>0}
{z^{-k}\over k}{\partial\over\partial t_k}}
\det\,(1-zA^{-1})\!: Z_{W_\infty}(t,\tbar).\cr} }
Inserting the eigenvalues $a_i$ of the matrix $A$, one can pick up the
residues of the poles at $z=a_i$, to get
\eqn\resid{
\eqalign{
{1\over \mu^2}{\partial Z_{W_\infty}\over \partial
\tbar_n}(t,\tbar) &=
{1\over n+1}:\! {1\over (i\mu)^{n+1}} \sum_i {a_i^{-i\mu}
{\rm e}^{{1\over i\mu}\sum_{k>0}
{a_i^{-k}\over k}{\partial\over\partial t_k}}\over
\prod_{j\ne i}(a_j - a_i)}\cr
& \qquad \times\left[\partial_z^{n+1} \big(z^{i\mu}
\prod_j (a_j-z)\,{\rm e}^{-{1\over i\mu}\sum_{k>0}
{z^{-k}\over k}{\partial\over\partial t_k}}
 \big)\right]_{z=a_i}\kern-2em : \kern0.8em Z_{W_\infty}(t,\tbar).\cr} }
Now, of the $n+1$ $z$-derivatives, at least one must act on the second
factor in the brackets, otherwise the term will vanish. So one
derivative can be picked out in $n+1$ ways to do this, cancelling the
$n+1$ factor in the denominator. Next, using the identity
\eqn\ident{
{\rm e}^{{1\over i\mu}\sum_{k>0}
{a_i^{-k}\over k}{\partial\over\partial t_k}}\;
\left[{\partial\over \partial
z} {\rm e}^{-{1\over i\mu}\sum_{k>0}
{z^{-k}\over k}{\partial\over\partial t_k}}\right]_{z=a_i}
 = {1\over i\mu} \sum_{k>0} a_i^{-k-1}{\partial\over\partial t_k}
= {\partial\over\partial a_i}, }
one rewrites Eq. \resid\ as
\eqn\residtwo{
\eqalign{
{1\over \mu^2}{\partial Z_{W_\infty}\over \partial
\tbar_n}(t,\tbar) &=
-{1\over (i\mu)^{n+1}} \sum_i {a_i^{-i\mu}\over
\prod_{j\ne i}(a_j - a_i)} \cr
& \qquad\qquad \times\left({\partial\over\partial a_i}\right)^n
\big(a_i^{i\mu}\prod_{j\ne i} (a_j-a_i) Z_{W_\infty}(t,\tbar)\big). \cr}}

Recalling the well-known result
\eqn\lapl{
\tr\left({\partial\over\partial A}\right)^n = {1\over \Delta(a)} \sum_i
\left({\partial\over \partial a_i}\right)^n \Delta(a),  }
where $\Delta(a)= \prod_{j<k}(a_j - a_k)$ is the Vandermonde
determinant, one can change back from eigenvalues to the full matrix $A$:
\eqn\remarkable{
{1\over i\mu}{\partial Z_{W_\infty}\over \partial
\tbar_n}(t,\tbar)
= {1\over (i\mu)^n} (\det\, A)^{-i\mu}\, {\rm
tr}\left({\partial\over\partial A}\right)^n (\det\,
A)^{i\mu}\, Z_{W_\infty}(t,\tbar). }
This is a remarkable expression for $W_\infty$ in terms of a single
fixed matrix!

This identity can be solved through random matrices in the
following way. Introduce the matrix integral
\eqn\intro{
Z_{K}(t,\tbar)  = (\det\,A)^{-i\mu} \int dM~ {\rm e}^{\tr\,V(M,t,\tbar)} }
with some, as yet unknown, potential $V$. Then Eq. \remarkable\ implies
that
\eqn\impl{
\left[ {1\over i\mu} {\partial\over\partial \tbar_n}- {1\over (i\mu)^n}
\tr\left(\partial\over\partial A\right)^n \right] \Big((\det\,A)^{i\mu}
Z_{K}(t,\tbar) \Big)=0. }
This determines
\eqn\determin{
V(M,t,\tbar) = i\mu\, MA + i\mu\,\sum_{k>0}\tbar_k M^k + f(M) }
where $f(M)$ is independent of $(t,\tbar)$.
The boundary condition is that $Z_{K}(t,0)$ must be independent of
$t$ (this comes from momentum conservation in the string
theory). Now
\eqn\tzero{
\eqalign{
Z_{K}(t,0) &=  (\det\,A)^{-i\mu} \int dM~ {\rm e}^{i\mu\,\tr\,MA +
\tr\,f(M)}\cr
&=  (\det\,A)^{-i\mu -N} \int dM~ {\rm e}^{i\mu\,\tr\,M +
\tr\,f(MA^{-1})},\cr}}
from which it follows that
\eqn\eff{
f(M) = -(i\mu +N) \log\,M }
and consequently also the matrix integral must be over positive-definite
Hermitian matrices $M$.

It follows that,
up to an overall multiplicative constant independent of $(t,\tbar)$,
$Z_{W_\infty}(t,\tbar)$
is equal to the Kontsevich-type integral
\eqn\kontstype{
Z_{K}(t,\tbar) = (\det\,A)^{-i\mu}\,\int dM~ {\rm e}^{i\mu\,
\tr M A - (i\mu + N) \tr\,\log M + i\mu\, \sum_{k>0}\, \tbar_k
\tr M^k}. }
This can equivalently be written
\eqn\equivalent{
Z_{K}(t,\tbar) = \int dM~ {\rm e}^{i\mu\,
\tr M - (i\mu + N) \tr\,\log M + i\mu\, \sum_{k>0}\, \tbar_k
\tr(MA^{-1})^k}. }
In this form, our model is very similar to the matrix models studied
recently by Kazakov et al.\ref\kazakov{\KAZAKOV}, with the difference
that the Gaussian potential is replaced by the gamma-function
integrand. The ``Euclidean'' continuation $\nu=-i\mu$ gives a
convergent integral as long as $\nu>N-1$.

\newsec{Previous Kontsevich-type models for $c=1$}
The ``Kontsevich-Penner'' model of Dijkgraaf et al. is given by the
following matrix integral:
\eqn\dmpmod{ Z_{DMP}(t,\tbar) = (\det\,A)^{-N+i\mu}
\int dM~ {\rm e}^{i\mu\, \tr [M A^{-1}
- \log M + \sum_{k>0}\, \tbar_k M^{-k}]}, }
with the parameters $t_n$ defined as in Eq. \frob\ above.

Comparing with Eq. \kontstype, we see that there are three differences:
(i) given the convention in Eq. \frob\ for the relation between $t_n$
and $A$, the power of $A$ appearing in the first term is negative in
the DMP model but positive in the correct one; (ii) the coefficient
of the $\log$ term is $-i\mu$ in the DMP model but $-i\mu -N$ in our
model; (iii) the perturbations representing the incoming tachyons are
negative powers of $M$ in the DMP model but positive powers in our
model.

These differences are in no way conventional or removable by any
change of variables. One of the simplest ways to see this is that in
our model there are {\it two} linear terms in $M$, one coupled to $A$
and the other to $\tbar_1$. This fact is responsible for the puncture
equation, as we show below. The DMP model has only one linear term in
$M$, and does not satisfy the puncture equation. It is quite
straightforward, if a little tedious, to compute correlators for the
DMP model using Schwinger-Dyson equations, and we will give some
examples below. By contrast, the Schwinger-Dyson equations are
virtually trivial for our model, perhaps the greatest surprise of the
present analysis.

As is well-known, Kontsevich-type matrix integrals make sense even for
$1\times 1$ matrices, where they compute correlators in some
1-dimensional subspace of the $(t,\tbar)$ parameter space. Thus, one
can start by computing $Z^{-1}
\partial Z/\partial\tbar_n$ at $\tbar=0$
in the DMP model of Eq. \dmpmod\ above, and for our Kontsevich-type
model (Eq. \kontstype), with the matrix $M$ replaced by a
single variable $m$, and with the constant matrix $A$ set equal to a
number $a$. This should be compared with the $W_\infty$ answer for the
same object, evaluated at $-i\mu\, t_n = {a^n\over n}$. The
calculations are elementary, and one finds
\eqn\onedimdmp{
\langle m^{-n} \rangle_{DMP} = \left({-i\mu\over a}\right)^n
{\Gamma(-i\mu -n +1)\over \Gamma(-i\mu +1)},}
while
\eqn\onedimkp{
\langle m^n \rangle_{K} = {1\over (i\mu a)^n}
{\Gamma(i\mu +1)\over \Gamma(i\mu -n +1)} }
and, from $W_\infty$,
\eqn\onedimw{
-i\mu\langle T_{-n} \rangle_{W_\infty} = {1\over (i\mu a)^n}
{\Gamma(i\mu +1)\over \Gamma(i\mu -n +1)}. }

Another useful way to compare the DMP model and ours
comes from the puncture equation. In the $W_\infty$ solution, the
simplest case $n=1$ leads to the recursion relation
\eqn\recur{
{\partial\cF \over \partial \tbar_1} = t_1 - (k+1)\,t_{k+1}
{\partial\cF \over \partial t_k}, }
where
\eqn\free{
Z_{W_\infty}(t,\tbar) = {\rm e}^{\mu^2 \cF}.}
Translated into the language of the Kontsevich-type model, this
implies that, for infinitesimal $\epsilon$, the partition
function should satisfy
\eqn\scalet{
Z(t_k + \epsilon (k+1)t_{k+1}, \tbar_1 + \delta_{k,1} \epsilon)
= {\rm e}^{\mu^2 \epsilon t_1} Z(t,\tbar).}
Now, the transformation on $Z$ in the LHS is equivalent to
transforming the couplings $\tbar_k$ and the constant matrix $A$ as
follows:
\eqn\scaleta{
\eqalign{\tbar_1 &\rightarrow \tbar_1 + \epsilon\cr
A &\rightarrow A - \epsilon .\cr} }
In the DMP model, this change does not lead to any Ward identity for
$Z$, as one can check, so Eq. \dmpmod\ does not satisfy the puncture
equation. However, in our model Eq. \kontstype, the variations of the
two linear terms in $M$ compensate each other under the transformation
Eq. \scaleta, leaving only a change from the determinant factor
outside, which precisely gives Eq. \scalet.

One more interesting point to observe is that both the DMP model and
ours can be rewritten after making the transformation $M\rightarrow
M^{-1}$, which is a legitimate change of variables since $M$ is a
positive matrix. One finds the alternative forms
\eqn\inverted{
\eqalign{
Z_{DMP}(t,\tbar) &= (\det\,A)^{-N+i\mu}
\int dM~ {\rm e}^{i\mu\, \tr [M^{-1} A^{-1} + (1-
{2N\over i\mu}) \log M + \sum_{k>0}\, \tbar_k M^{k}]}\cr
Z_{K}(t,\tbar) &= (\det\,A)^{-i\mu}\,\int dM~ {\rm e}^{i\mu\,
\tr M^{-1} A - (-i\mu + N) \tr\,\log M + i\mu\, \sum_{k>0}\,
\tbar_k\tr M^{-k}}.\cr } }
In the DMP model, this inversion introduces an explicit $N$ into the
potential, which was not there before. However, in our model which
already contained an $N$ factor, it reappears in the same form, due
quite simply to the change of matrix measure under inversion. Indeed,
our model in the singular limit $\mu \rightarrow 0$ is manifestly
symmetric under inversion, as it becomes the matrix analogue of $\int
dx/x$.

Finally, we present the results of explicit computations of a class of
simple correlation functions for the DMP model, and for our model, to
all orders in ${1\over\mu}$.  The former are obtained using
Schwinger-Dyson equations in the form
\eqn\schw{
0=\int dM~ {
\partial\over\partial M_{ij}}\left({\rm e}^{i\mu\, \tr [M A^{-1} -
\log M]}\, f_{ij}(M,A)\right), }
where $f_{ij}(M,A)$ is an arbitrary matrix-valued function.
This process is rather tedious, particularly for the last line of the
table, since one has to write down the above equation for a large
number of choices of the function $f_{ij}$ and then successively
eliminate terms to get the desired result. In contrast, the relevant
Schwinger-Dyson equations for our model follow from Eq. \remarkable
\foot{In collaboration with V. Kazakov, we have
also carried out these computations using the technique of character
expansions\kazakov, applied to the version \equivalent\ of our model
and the analogous one for the DMP model, and we obtained the same
results. In particular, in this formalism it is manifest that both
models are symmetric under exchange of $t$ and $\tbar$. Character
expansion turns out to be the most efficient technique for the DMP
model.}.

$$
\vbox{\tabskip=0pt
\setbox\strutbox=\hbox{\vrule height12pt depth8pt width0pt}
\halign{\strut#& \vrule#& \hfil#\hfil &
\vrule#& \hfil#\hfil&\vrule#& \hfil#\hfil & \vrule#\tabskip=0pt\cr
\noalign{\hrule}
& & ~Correlator & & $c=1$ Kontsevich model & & DMP model
&\cr\noalign{\hrule}
& & $\langle T_{-1} \rangle$ & & $t_1$ & & $t_1$ &\cr\noalign{\hrule}
& & $\langle T_{-2} \rangle$ & & $2t_2 + t_1^2$ & &
${\mu^2\over 1+\mu^2} (2t_2 + t_1^2)$ &\cr\noalign{\hrule}
& & $\langle T_{-3} \rangle$ & & $3t_3 + 6 t_1 t_2 +
t_1^3 + {1\over(i\mu)^2}3t_3$
& & ${\mu^4\over (1+\mu^2)(4+\mu^2)}(3t_3 + 6 t_1 t_2 +
2t_1^3)$ &\cr\noalign{\hrule}
& & $\langle T_{-4} \rangle$ \phantom{\vrule height0pt depth40pt width0pt}
 & & \vtop{\hbox{$~4t_4 + 12 t_1 t_3 +
8t_2^2 + 12 t_1^2 t_2 + t_1^4 $}\hbox{$\quad +
{1\over(i\mu)^2}(20 t_4 + 4t_2^2 + 12 t_1 t_3)$}} & &
\vtop{\hbox{${\mu^6\over (1+\mu^2)(4+\mu^2)(9+\mu^2)}
\big(4t_4 + 12 t_1 t_3$} \hbox{$\qquad\qquad ~
+ 8 t_2^2 + 20 t_1^2 t_2 + 5 t_1^4$}
\hbox{$ \quad  +
{1\over(i\mu)^2}(4t_4 -12 t_2^2 + 12t_1 t_3)
\big)$}} &\cr\noalign{\hrule}
}}
$$
\nobreak
\centerline{Table I: Explicit computations of amplitudes}
\bigskip
It is easy to see from this that the DMP model is inequivalent to our
model, and to $W_\infty$.

Let us briefly mention that Chekhov and Makeenko\chekmak\ had proposed
a model which they conjectured to be in some sense equivalent to the
$c=1$ string, though they did not state precisely what the full
equivalence should be. Their matrix potential was
\eqn\chek{
V(M,A) = N\Big(MA + \nu\log M - \half M^2\Big) }
with $t_k = {1\over k}\tr A^{-k} - {N\over 2}\delta_{k,2}$ and with
$\nu$ being the cosmological constant. This has some similarities to
Eq. \kontstype, but the dependence on $\nu$ and $N$ is not the correct
one.

\newsec{The derivation via semi-infinite forms}
In Ref.\dmp\ a ``coherent state'' representation for the generating
functional $Z(t,\tbar)$ was derived. This representation involves a
Fock space associated to bosonic creation and annihilation operators
$\amn$ and $\an$, satisfying the canonical commutation relations
$\left[\alpha_m ,\an\right] = m\delta_{m+n,0}$ with $m, n =1,2,\ldots$
The $\an$ are conveniently collected into the conformal current
$\partial\varphi (z) \equiv \sum_n \an z^{-n-1}$, which is related to
the fermionic fields
\eqn\ffields{
\psi(z) = \sum_{n\in \ZZ} \psi_{n+\half} z^{-n-1}\qquad\qquad
\psibar (z) = \sum_{n\in \ZZ} \psibar_{n+\half} z^{-n-1},}
by the familiar 2-dimensional bosonization formulas: $\partial\varphi(z)
= ~:\!\psibar(z)\psi(z)\!:~$. The fermionic oscillators in Eq.
\ffields\ obey canonical anticommutation relations:
$\lbrace \psi_r ,\psibar_s\rbrace = \delta_{r+s,0}$, with
$r,s \in \ZZ +\half$.

The coherent state formula of Dijkgraaf et al. for the partition function
of 2D string theory is
\eqn\cstates{
 Z ( t, \tbar) = \langle t | S | \tbar\rangle ,}
where $\langle t|$ and $| \tbar\rangle$ are coherent
states associated to the positive and negative tachyons:
\eqn\tcoherent{
\langle t | \equiv \langle 0| {\rm e}^{ i\mu
\sum_{n=1}^\infty \an t_n}\equiv \langle 0| U(t)\qquad\qquad
|\tbar \rangle \equiv {\rm e}^{ i\mu
\sum_{n=1}^\infty \amn \tbar_n} |0\rangle \equiv U(\tbar)|0\rangle .}
The operator $S$ acts linearly on the fermionic fields:
\eqn\saction{
S \psi_{-n-\half} S^{-1} = R_{p_n} \psi_{-n-\half}
\qquad\qquad S \psibar_{-n-\half} S^{-1} = R^*_{p_n}
\psibar_{-n-\half},}
where $R_{p_n}$ are reflection coefficients depending on the fermionic
momentum $p_n = n+\half$ and satisfying the unitarity condition
$R_{p_n} R^*_{-p_n} = 1$.

The matrix model formulation of the $c=1$ string theory leads to
explicit expressions for the reflection coefficients\mooretc,\dmp:
\eqn\reflection{
R_{p_n} = (-i\mu)^{-p_n}{\Gamma (\half -i\mu + p_n)\over
\Gamma (\half-i\mu)}.}

The strategy to derive a Konsevitch model from the coherent states
formula \cstates\ is to represent the fermionic
Fock space in terms of semi-infinite forms.
Let us make the same choice as in Ref.\dmp\ for the semi-infinite form
representing the fermionic Fock vacuum:
\eqn\fvacuum{
|0\rangle = z^0 \wedge z^1\wedge z^2 \ldots }
Now we must take representatives for $\psi_{n+\half}$ and
$\psibar_{n+\half}$
which, for $n >0$, {\it annihilate} the vacuum. This is necessary since
the coherent state formula \cstates\ and the action of the
$S$ operator \saction\ are defined assuming such a convention,
which is also the standard one in conformal field theory.
A representation consistent with this convention is
\eqn\faction{
\psi_{n+\half} = z^n , \qquad \psibar_{-n-\half}=
{\partial\over \partial z^n}.}
It follows from Eq. \saction\ that
\eqn\semisaction{
S: z^n \rightarrow  R_{-p_n}z^n .}
The error in the derivation of the Konsevitch-Penner model of
Ref. \dmp\ is precisely a choice of representatives for the
fermionic operators which is inconsistent with that of conformal field
theory. With their choice, the action of $S$ was $z^n \rightarrow
R_{p_n}z^n$ (Eq. (5.26) of their paper).

In the following we briefly trace back the steps of
the derivation of the Konsevitch model for $c=1$ which starts
from the coherent state formula, and show that, once the correct
choice \faction\ is made, one recovers our matrix model.

Recalling that $\bigl[\an , \psi_{m+\half}\bigr] = \psi_{m+n+\half}$,
the action of the coherent state operator $U(\tbar)$ on the fermionic
oscillators
\eqn\uaction{
U(\tbar) :
\psi_{n+\half} \rightarrow U(\tbar) \psi_{n+\half} U(\tbar)^{-1}}
reads in the semi-infinite forms representation as follows
\eqn\uexplicit{
\eqalign{
U(\tbar): z^n &\rightarrow {\rm e}^{i\mu \sum_{k>0}
\tbar_k \alpha_{-k}} z^n
{\rm e}^{-i\mu \sum_{k>0} \tbar_k \alpha_{-k}} \cr
&= {\rm e}^{i\mu \sum_{k>0} \tbar_k z^{-k}}z^n = \sum_{k=0}^{\infty}
P_k (i\mu\tbar) z^{n-k},\cr}}
where the $P_k(i\mu\tbar)$ are the Schur polynomials.

Therefore the combined action of $S$ and $U(\tbar)$ is
\eqn\suaction{
\eqalign{
S \circ U(\tbar): z^n \rightarrow\, w^{(n)}(z;\tbar)&=
S \sum_{k=0}^{\infty} P_k (i\mu\tbar) z^{n-k}S^{-1}\cr
&= \sum_{k=0}^{\infty} P_k(i\mu\tbar) R_{-p_{n-k}} z^{n-k}.\cr}}

Recalling the expression \reflection\ for the reflection coefficients
and rewriting the gamma-function in terms of its integral representation
one obtains
\eqn\suactionbis{
\eqalign{
w^{(n)}(z;\tbar) &=
{(-i\mu)^{\half }\over \Gamma (\half -i\mu)}
\int_0^{\infty} dm~{\rm e}^{-m} m^{-i\mu-1}
\sum_{k=0}^{\infty} P_k(i\mu\tbar)
\left({-i\mu z\over m}\right)^{n-k}\cr
&= \;c(\mu)\, z^{-i\mu}
\int_0^{\infty} dm~ m^{-n}{\rm e}^{i\mu z m} m^{-i\mu -1}
{\rm e}^{i\mu \sum_{k>0}\,\tbar_k m^k},\cr}}
where
\eqn\cofmu{
c(\mu) \equiv {(-i\mu)^{-i\mu
+ \half}\over \Gamma(\half -i\mu)}.}

{}From this we finally derive the expression for the state $S |\tbar
\rangle$ in terms of semi-infinite forms
\eqn\suvacuum{
\eqalign{
S |\tbar \rangle =&\; S \circ U(\tbar)\, z^0 \wedge z^1 \wedge
z^2\wedge \ldots \cr
=&\; w^{(0)}(z;\tbar)\wedge w^{(1)}(z;\tbar)\wedge w^{(2)}
(z;\tbar)\wedge \ldots\cr}}

One also needs to make use of the
parametrization \frob\ for the coherent state $\langle t|$. If $a_i$,
with $i=1,\ldots, N$ are the eigenvalues of the Hermitian matrix $A$
in Eq. \frob , then
\eqn\comiwa{
\langle t | = \langle 0 | \prod_{i=1}^N {\rm
e}^{-\sum_{n>0} {\an\over n} a_i^n} =
\langle N |{ \prod_{i=1}^N \psi (a_i)\over \Delta(a)},}
where the state $| N \rangle$ reads as follows in the
semi-infinite form representation:
\eqn\nstate{
|N\rangle = z^N \wedge z^{N+1}\wedge z^{N+2}\ldots}
Putting together the bra in Eq. \comiwa\ with the ket in Eq. \suvacuum\
one gets the formula expressing $Z(t ,\tbar)$ in terms of determinants:
\eqn\zdeterm{
\eqalign{
Z(t,\tbar) &= \langle t | S |\tbar\rangle = {\det\,
w^{(j-1)}(a_i)\over \Delta(a)}\cr
&= c(\mu)^N (\prod_j a_j)^{-i\mu} \cr
&\qquad\times \int_0^\infty
\prod_j \left({dm_j\over m_j}~
{\rm e}^{i\mu m_j a_j -i\mu \log m_j + i\mu
\sum_{k>0}\, \tbar_k m_j^k}\right)
{\Delta(m^{-1})\over\Delta(a)}.\cr} }

Converting the Vandermonde depending on $m_i^{-1}$ to the standard one,
and using the Harish-Chandra formula, one finds (up to overall factors
independent of $(t,\tbar)$):
\eqn\harish{
\eqalign{
Z(t,\tbar)&= \Big(\prod_j a_j\Big)^{-i\mu}
\int_0^\infty \prod_j \left({dm_j\over m_j^N}~
{\rm e}^{i\mu m_j a_j -i\mu \log m_j + i\mu \sum_{k>0}\, \tbar_k m_j^k}\right)
{\Delta(m)\over\Delta(a)}\cr
&= (\det\,A)^{-i\mu} \int dM~ {\rm e}^{i\mu\,
\tr M A - (i\mu + N) \tr\,\log M + i\mu\, \sum_{k>0}\, \tbar_k
\tr M^k},\cr}}
which is precisely our model, Eq. \kontstype.

\newsec{Relation to Penner and Kontsevich models}
Let us set the couplings $t_k=\tbar_k=0$ in Eq. \kontstype. Then we are
left with a partition function
\eqn\teqzero{
Z(\mu,N) = \int dM~ {\rm e}^{i\mu\tr M - (i\mu +N)\tr\,\log M}. }
Rescaling and shifting $M$, we find
\eqn\resc{
Z(\mu,N) = {\rm e}^{N(i\mu +N)}\left(1+{i\mu\over N}\right)^N
\int dM~ {\rm e}^{(i\mu+N)\sum_{k=2}^\infty \tr {M^k\over k} },}
which is proportional to the Penner integral
\eqn\pennerint{
Z_{Penner}(\mu,N) = \int dM~ {\rm e}^{-Nt\sum_{k=2}^\infty
\tr {M^k\over k}} }
where $t=-(1+{i\mu\over N})$.

This integral was devised by Penner to count the Euler characters
$\chi_{g,n}$ of Riemann surfaces with genus $g$ and $n$ punctures.
However, as Distler and Vafa\ref\distvaf{\DISTVAF} showed, the double
scaling limit $N\rightarrow\infty$ and $t\rightarrow t_c=-1$ with
$\nu = N(t-t_c)$ fixed, actually counts the Euler characteristic
of {\it unpunctured} surfaces. Clearly, in Eq. \pennerint\ above this
is just the limit $N\rightarrow\infty$ with $\mu$ fixed, and we have
$\nu=-i\mu$ as the relation between the cosmological constant of
\distvaf\ and ours. So the limit that is in the spirit of Kontsevich
integrals is quite the same as Distler and Vafa's double-scaling
limit, contrary to the claim in Ref.\dmp.

Let us now examine some other limits of our model. To start with, it
is convenient to Euclideanize the cosmological constant via
$\nu=-i\mu$. Next, setting $\nu=N$ and $\tbar_k = \delta_{k,3}$ one
has:
\eqn\kontsorg{ Z_{K}(\nu=N, t, \tbar=\delta_{k,3}) =
(\det\,A)^N \,\int dM~ {\rm e}^{-N\,\tr M A -N\tr M^3}. }
The $\log$ term and the expansion parameter $\nu$ have both
disappeared simultaneously, and we have obtained a matrix integral
which is very similar to the original Kontsevich model describing
intersection theory on the moduli space of Riemann surfaces. Indeed,
the above integral would be the matrix Airy integral but for the fact
that integration is performed over positive-definite
matrices. However, the asymptotic expansion of this integral, based as
it is at the saddle-point $M\sim\sqrt{A}$, does not see this
difference. Indeed, Kontsevich shows\konts\ that the matrix Airy
integral gives rise to a sum over $2^N$ Kontsevich matrix models. In
contrast, the integral in Eq. \kontsorg\ above satisfies an
inhomogeneous version of the Airy equation because of the boundary of
the integration region at 0. It has a unique saddle-point by virtue
of the positive-definiteness of $M$, so that up to the usual factors,
it leads to precisely one Kontsevich model.

Thus the original Kontsevich model of two-dimensional pure gravity can
be thought of as a special case of our $c=1$ Kontsevich-type model
(but not of the DMP model), after some suitable scalings and
normalizations.  The same is true for the generalized Kontsevich
models, which appear by setting $\tbar_k = \delta_{k,p+1}$ for some
$p>2$. Note that in this picture, the choice of a fixed $\tbar_k$ and
$t_k$ will ultimately correspond to a choice of $(p,q)$ specifying a
definite $c<1$ minimal model coupled to gravity. The symmetry of the
$(p,q)$ minimal models in $p$ and $q$ would then be due to the
symmetry of the $c=1$ theory in $t_k$ and $\tbar_k$.
We will comment further on the significance of these
points below.

Since the Kontsevich and generalized Kontsevich models are special
cases of our model, it should follow that the Virasoro and $W_n$
identities satisfied by the former arise from the $W_\infty$ of the
latter. This does not imply a completely straightforward connection,
however, since the passage to Kontsevich models requires several
rescalings and normalization factors. Additionally, the couplings of
the Kontsevich model are defined in terms of the matrix $A$ not
through Eq.\frob, but rather through a twisted version of it: $t_k\sim
\tr A^{-k-\half}$, the shift by $\half$ being responsible for the
``twisted free bosons'' investigated in Refs.\vir,\ref\im{\IM}.
Similar fractional shifts occur for the generalized Kontsevich models.

\newsec{Quantum Effective Action}
We have seen in  Section 2 that correlators of negative tachyons $T_{-n}$
of $c=1$ string at $\tbar=0$ are equal to the averages of
$\tr\, {M^n\over \nu}$ taken with the matrix measure:
\eqn\measure{\eqalign{
Z_{\Gamma}(A) &= \int dM~ {\rm e}^{-\nu\,\tr M A +(\nu -N) \tr\log
M}\cr  &= (\det\,A )^{-\nu}
\int dM~ {\rm e}^{-\nu\,\tr M +(\nu -N) \tr\log M}.\cr}}
The matrix integral above has the obvious property that adding an
external source $J$ for $M$ in the classical action leaves the form
of the integrand invariant:
\eqn\measuresource{\eqalign{
Z_{\Gamma}(A;J)
&= \int dM~ {\rm e}^{-\nu\,\tr\, M A +(\nu -N) \tr\log
M -\tr\, J M}\cr  &= \left(\det\, (A +{J\over \nu} )\right)^{-\nu}
\int dM~ {\rm e}^{-\nu\,\tr\, M +(\nu -N) \tr\log M}.\cr}}
Let us define the free energy to be minus the log of this expression,
dropping the additive constant coming from the integral. Thus:
\eqn\fre{
F_A(J) = \nu\,\tr\log\left(A + {J\over\nu}\right). }
This allows us to derive explicitly the quantum action associated to
this matrix measure. Define the ``quantum'' field $\Mhat$ via
the equation
\eqn\qmatr{\Mhat = {\partial F_A(J)\over \partial J}= \left( A +
{J\over \nu}\right)^{-1}.}
The quantum action $\Gamma (\Mhat)$ for $\Mhat$ is defined through
a Legendre transformation of $F_A(J)$
\eqn\qaction{\Gamma (\Mhat) = F_A(J) - \tr\, \Mhat J,}
and can be easily evaluated to give
\eqn\qactioneva{\Gamma (\Mhat) = -\nu\, N +\nu\, \tr\, \Mhat A
-\nu\,\tr\log\Mhat .}
The form of the quantum action is identical to that of the
``classical'' action in Eq. \measure , the only difference being
two simple renormalization effects: the appearance of a
constant zero-point energy and the renormalization of the
coefficient of the logarithm, which becomes $N$-independent.

The renormalization of the log term is extremely important. Even if we
had set $\nu=N$ in Eq. \measure\ and thereby eliminated the log term
in the classical action, it would still be present in the quantum
action --- thus it is {\it dynamically generated}. It cannot be tuned
away as long as the background has all $\tbar_k=0$.

The quantum action leads to the equation of motion
\eqn\emotion{0 ={\partial \Gamma(\Mhat)\over\partial\Mhat}=
A - \Mhat^{-1},}
i.e. $\langle \Mhat \rangle = A^{-1}$. This means that the quantum
field $\Mhat$ has to be shifted around its vacuum expectation
value,
\eqn\shfield{\Mhat = A^{-1} + \mhat ,}
and the quantum action becomes
\eqn\qpenaction{
\eqalign{
\Gamma (\Mhat) &= \nu\, \tr \log A + \nu\,\tr A\mhat
-\nu\,\tr\log (1 + A\mhat)\cr
&= \nu\,\tr\log A  +\nu\,\sum_{k=2}^{\infty} {(-1)^k\over k}\tr
(A\mhat)^k.\cr}}
This expression, which  encodes the full perturbation series for tachyon
scattering in $c=1$ string theory, might be called the Penner quantum action.
It corresponds to our Kontsevich-type model shifted around the
classical solution appropriate to $c=1$ string theory.

The 1PI vertices for the field $\mhat$ can be read off from
Eq. \qpenaction; for example the 2-point and 3-point vertices are
\eqn\nvertices{
\eqalign{
\Gamma^{(2)}_{i_1j_1;i_2j_2} &= \nu\,A_{i_2j_1}A_{i_1j_2}\cr
\Gamma^{(3)}_{i_1j_1;i_2j_2;i_3j_3}&=
{\half}\nu\,\left[ A_{i_3j_1}A_{i_1j_2}A_{i_2j_3} +
A_{i_2j_1}A_{i_3j_2}A_{i_1j_3}\right].\cr}}
Tree diagrams built out of these 1PI $n$-point vertices
$\Gamma_{i_1j_1;\ldots ;i_nj_n}$ together with the exact propagator
$G^{(2)}_{i_1j_1;i_2j_2} = \langle \mhat_{i_1j_1} \mhat_{i_2j_2}\rangle =
{1\over \nu} A^{-1}_{i_1j_2} A^{-1}_{i_2j_1}$ generate all
correlators $\langle M_{i_1j_1;\ldots ;i_nj_n}\rangle$ and therefore
reproduce all negative-tachyon expectation values. For example,
$\nu \langle T_{-2}\rangle $ is given by
\eqn\twoaverage{ \langle \tr\, \Mhat^2\rangle = \tr\,A^{-2} + \langle
\tr\,\mhat^2\rangle = \tr\, A^{-2} + {1\over\nu}(\tr\, A^{-1})^2,}
while for $\nu \langle T_{-3}\rangle$ one obtains,
\eqn\threeaverage{
\eqalign{
\langle \tr\, \Mhat^3\rangle &= \tr\,A^{-3} + \langle\tr\, A\mhat^2
\rangle + \langle\tr\, \mhat^3\rangle \cr
&= \tr\, A^{-3} + {1\over\nu}\tr\, A^{-2}\tr\,A^{-1} + {1\over\nu^2}
\left( \tr\, A^{-3}+ ( \tr\, A^{-1})^3\right),\cr}}
in agreement with the results shown in the table of Section 3.

To summarise, the polynomials in $t$ given by the negative-tachyon
correlators of $c=1$ string admit a neat diagrammatical
interpretation, as a sum over connected and disconnected tree diagrams
of the quantum Penner action \qpenaction .

\newsec{Background Independence}
Suppose that one tried to build up the most trivial matrix model
possible, with a single Hermitian {\it positive-definite} matrix
$M$. One might imagine choosing the potential to be zero, and then
introducing an external source $A$:
\eqn\triv{
Z(A) = \int dM~{\rm e}^{-\nu\,\tr MA}.}
{}From Eq. \qactioneva\ it follows that the quantum effective action,
up to additive constants, is
\eqn\qactiontwo{\Gamma (\Mhat) = \nu\, \tr \Mhat A
-N\,\tr\log\Mhat.}
Since a logarithmic term has appeared from renormalization effects, it
is natural to add a ``bare'' log term in the original action. Choosing
the coefficient of this term so that the quantum action becomes
$N$-independent, we find:
\eqn\ourmodel{
Z(A) = \int dM~{\rm e}^{-\nu\,\tr MA + (\nu-N)\tr\,\log M}.}
This is precisely our Kontsevich-type model! Viewed as a string field
theory, the quantum equation of motion tells us that $\Mhat=A^{-1}$,
and expanding around this gives rise to the quantum Penner action that
we have already discussed. Therefore the $c=1$ string in this
framework is nothing but a positive-definite matrix with zero
potential, coupled to an external source.

What about other non-critical string backgrounds? Let us add the term
$\tr M^{k+3}$, for some fixed $k\ge 0$, to the above potential. This
corresponds to turning on a source for the tachyon $T_{-k-3}$ in $c=1$
language. One can no longer explicitly compute the quantum
action. However, the matrix integral now has a saddle-point which is
very far from $M=A^{-1}$. Indeed, tuning away the log term, the
saddle-point is at $M\sim A^{1\over k+2}$. Expanding around this
saddle-point leads to the generalized Kontsevich model of level
$k$~\konts\genkonts, which describes the $(k+2,q)$ minimal-model string
backgrounds. In this sense, all the $c<1$ string backgrounds can be
thought of as the different vacua to which our Kontsevich-type model
flows when there is ``condensation of negative tachyons''.

This picture needs to be studied in more detail, in particular to
understand what is the mechanism by which the log term gets tuned
away. Note that switching on $T_{-1}$ does not shift the saddle-point
since it can be absorbed in the source $A$, while $T_{-2}$ leads to a
background in which the free energy is quadratic in $A$ and hence
trivial.

\newsec{Conclusions}
We have solved the $W_\infty$ constraints of $c=1$ string theory via a
Kontsevich-type matrix model. The resulting model is beautiful and
natural, and we believe it should tell us something fundamental about
string theory. In particular, this could lead to a framework to
formulate background-independence in string field theory.

On the way, we obtained an elegant matrix version of the $W_\infty$
constraints, Eq. \remarkable, which has the form of a generalized
heat-kernel equation, Eq. \impl. One may be tempted to speculate that
this is related to the holomorphic anomaly equation of
Ref.\ref\bcov{\BCOV}\ which, according to Ref.\ref\wittenbk{\WITTENBK}\
expresses quantum background-independence in certain solvable
topological string theories.

Our results also shed new light on the sense in which $c=1$ string
theory is like a $k\rightarrow -3$ limit of the $k$-minimal
topological models coupled to gravity\ref\kminus{\KMINUS}.

It should be emphasized that our model was constructed starting from
the tachyon $S$-matrix and, apparently, does not contain the
other kinds of states that one might expect to see in two-dimensional
string theory, including the discrete ``tensor'' states and the states
of the ``wrong dressing''. However, our matrix-model in principle
contains many more operators than the ones we have considered, in
particular traces of negative powers, $\tr M^{-k}$ and also more
complicated objects such as $\tr (M^{k_1}A^{k_2}M^{k_3}\ldots)$. It
remains to be seen whether these provide the missing states of $c=1$.

Because of the resemblance of this model to those studied recently by
Kazakov et~al.\kazakov, we expect that the powerful technique of
character expansions can be used to gain more understanding of our
model and its possible generalizations.
\bigskip
\noindent{\bf Acknowledgements}

{\hyphenpenalty=10000
We are indebted to Volodya Kazakov for explaining to us the results of
Ref.\kazakov\ and for several helpful comments and suggestions. We
thank Robbert Dijkgraaf, Rajesh Gopakumar and Ivan Kostov for useful
discussions. Part of this work was carried out at the ICTP Trieste, whose
hospitality we gratefully acknowledge.

}

\listrefs
\bye